\documentclass[useAMS,usenatbib]{mn2e}

\usepackage{graphicx,psfig}

\title[Ultra-compact structure in J1819$+$3845]{Emergence and disappearance of micro-arcsecond structure in the scintillating quasar J1819$+$3845}
\author[J.-P. Macquart and A.G. de Bruyn]{J.-P. Macquart$^{1}$\thanks{E-mail:
jpm@astro.caltech.edu}\thanks{NRAO Jansky Fellow} and A. G. de Bruyn$^{2}$\thanks{Email: ger@astron.nl}\\
$^{1}$National Radio Astronomy Observatory, P.O. Box 0, Socorro NM 87801, U.S.A. and \\
Astronomy Department, Mail Code 105-24, California Institute of Technology, Pasadena CA 91125, U.S.A. \\
$^{2}$ Netherlands Foundation for Research in Astronomy, Dwingeloo, The Netherlands and \\ Kapteyn Astronomical Institute, University of Groningen, P.O. Box 800, Groningen 9700 AV, The Netherlands}
\begin{document}

\date{accepted}

\pagerange{\pageref{firstpage}--\pageref{lastpage}} \pubyear{2006}

\maketitle

\label{firstpage}

\begin{abstract}
The 4.8\,GHz lightcurves of the scintillating intra-day variable quasar J1819$+$3845 during 2004-5 exhibit sharp structure, down to a time scale of 15\,minutes, that was absent from lightcurves taken prior to this period and from the 2006 lightcurves.  Analysis of the lightcurve power spectra show that the variations must be due to the emergence of new structure in the source.  The power spectra yield a scattering screen distance of $3.8 \pm 0.3$\,pc for a best-fit $v_{\rm ISS}=59\pm 0.5\,$km\,$s^{-1}$ or $2.0 \pm 0.3\,$pc for the scintillation velocity reported by Dennett-Thorpe \& de Bruyn (2003).  The turbulence is required to be exceptionally turbulent, with $C_N^2 \ga 0.7 \, \Delta L_{\rm pc}^{-1} \,$m$^{-20/3}$ for scattering material of thickness $\Delta L_{\rm pc}$\,pc along the ray path.  The 2004 power spectrum can be explained in terms of a double source with a component separation $240 \pm 15\,\mu$as in 2004.
\end{abstract}


\section{Introduction}

The quasar J1819$+$3845 exhibits 20--35\% rms intensity modulations on 
intra-hour timescales at centimetre wavelengths.  The variations are due 
to interstellar scintillation caused by plasma thought to be 
located only $z=$4--12\,pc from Earth (Dennett-Thorpe \& de Bruyn 2002, 2003, 
hereafter DB03). For $z=$10\,pc the $\sim\! 30\,$min 
scintillations observed in the period December to March each year imply an 
overall source size of $\sim\!60\,\mu$as, corresponding to 0.39\,pc at 
its redshift of 0.54 ($H_0 = 70\,$km\,s$^{-1}$Mpc$^{-1}$, 
$\Omega_M=0.27$ and $\Omega_\Lambda = 0.73$).

The source has exhibited variations every time since they were first observed in 1999.  This is surprising 
because it might be supposed that such a compact source, with a brightness 
temperature exceeding the inverse Compton limit, should expand and cease 
scintillation after several months.  This problem is common to several 
persistent extreme intra-day variable sources including PKS\,1519$-$273 and 
PKS\,1257$-$326 (Macquart et al. 2000, Bignall et al. 2003).  At the very least, one would expect such compact sources to exhibit some degree of 
structural variability, made evident by changes in the variability 
characteristics.

The detection of structural changes in intra-hour variable sources is 
complicated, however, by annual cycles in their variability time scale 
(DB03; Bignall et al. 2003).  Both the amplitude and direction of the 
scintillation velocity change as the Earth's velocity changes relative to 
the scattering medium as it orbits the Sun.  This changes the 
scintillation direction, which in turn alters the character of the 
scintillations (i.e. its power spectrum) if the turbulence in the 
scattering medium is anisotropic or if the source structure is asymmetric.  
Both asymmetries hinder searches for structural variability.  These 
asymmetries are particularly important for J1819$+$3845 because its 
scintillation pattern is anisotropic, with an axial ratio of 
$15^{+>30}_{-8}$ (DB03).

In this paper we identify the emergence of a new structure within 
J1819$+$3845 at $\lambda 6$\,cm by comparing observations taken at 
identical epochs each year, thus removing ambiguity introduced by the annual cycle between intrinsic 
source structural evolution and asymmetry in the scintillation pattern.  In \S2 we present the lightcurves 
indicating the emergence of this new structure.  In \S3 the lightcurve 
power spectra are analysed to deduce its properties.  The implications of 
the discovery are summarised in \S4.

\section{Observations}

\begin{figure*}
\begin{center}
 \begin{tabular}{ccc}
\psfig{file=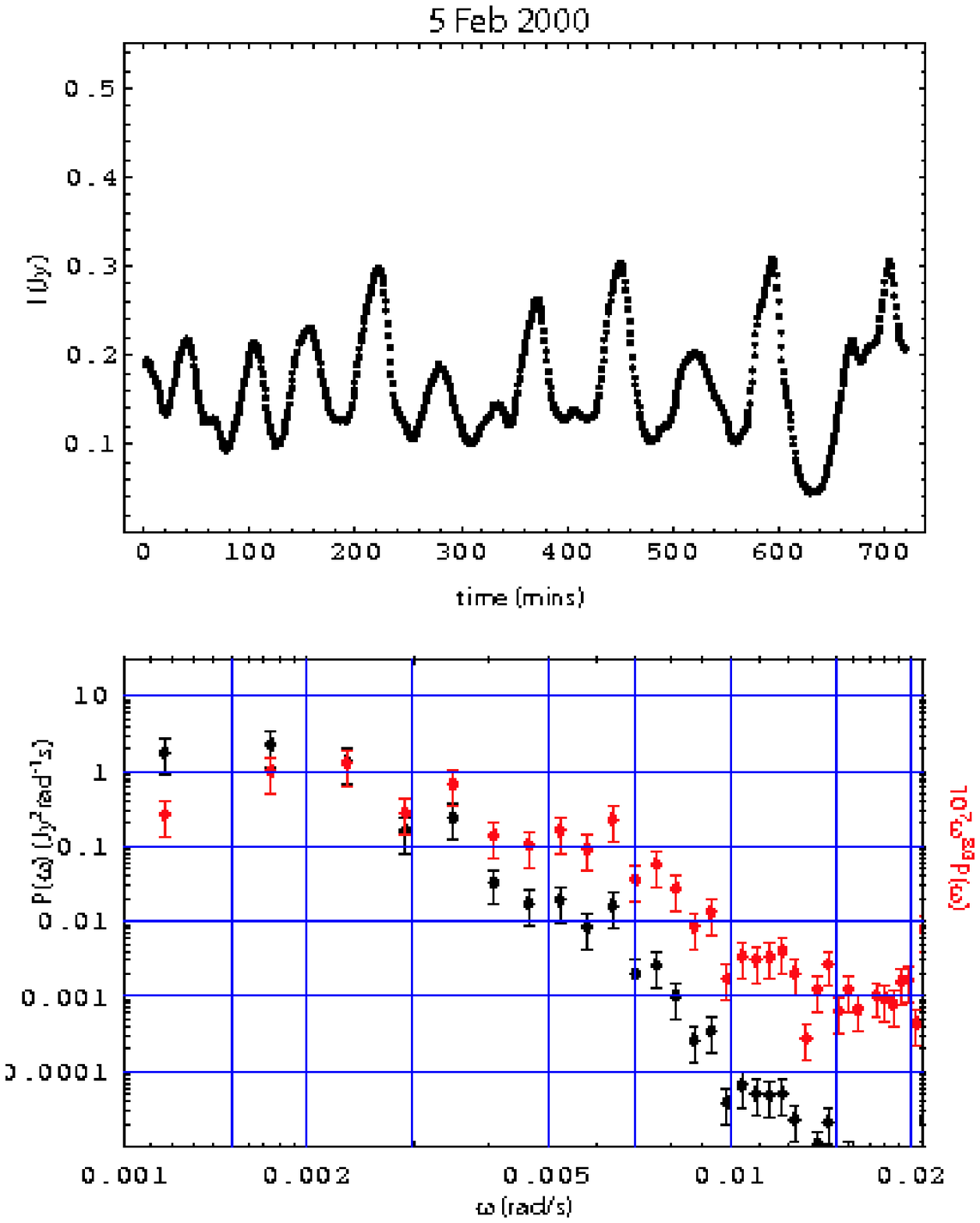,width=6.0cm} & \psfig{file=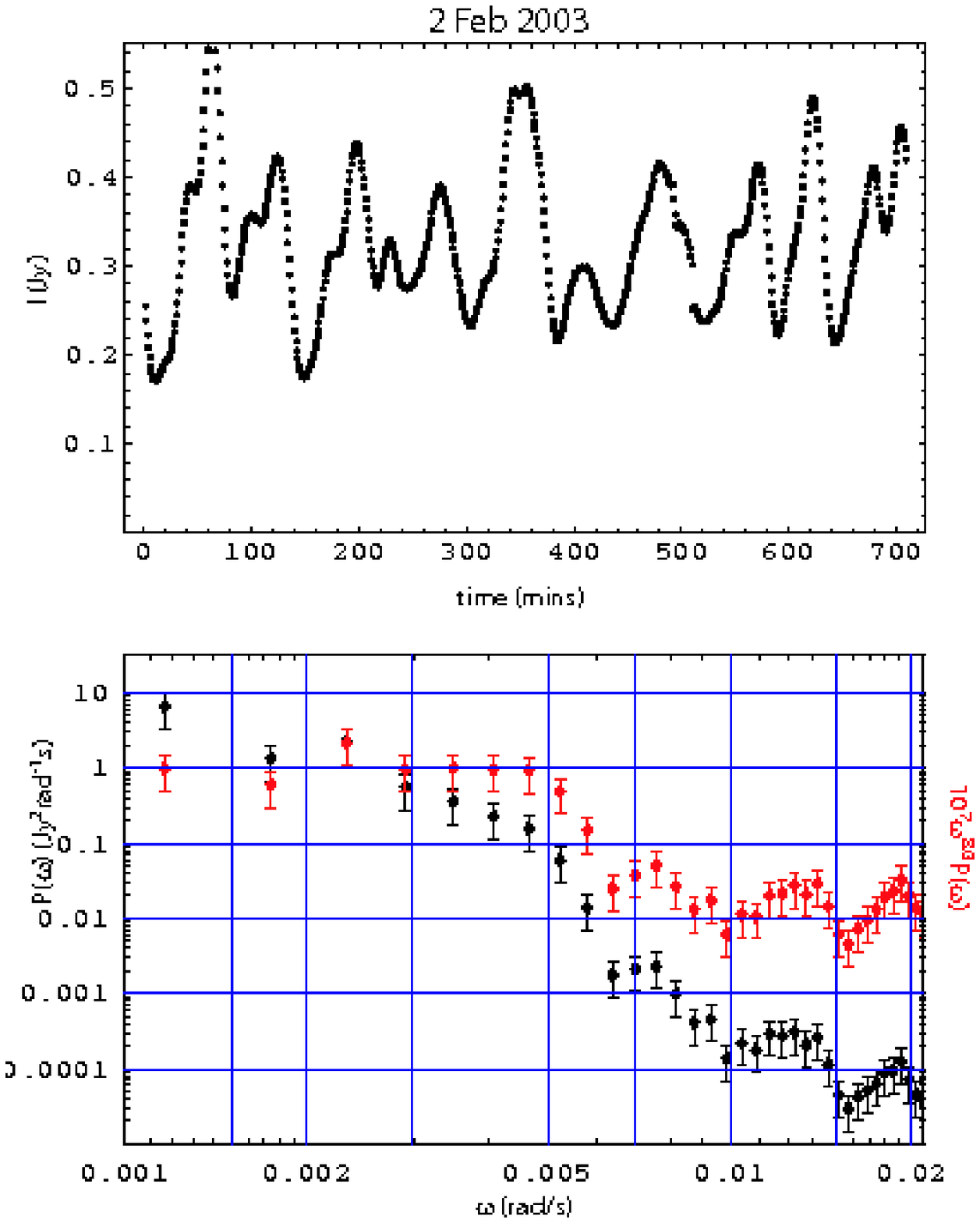,width=6.0cm} & \psfig{file=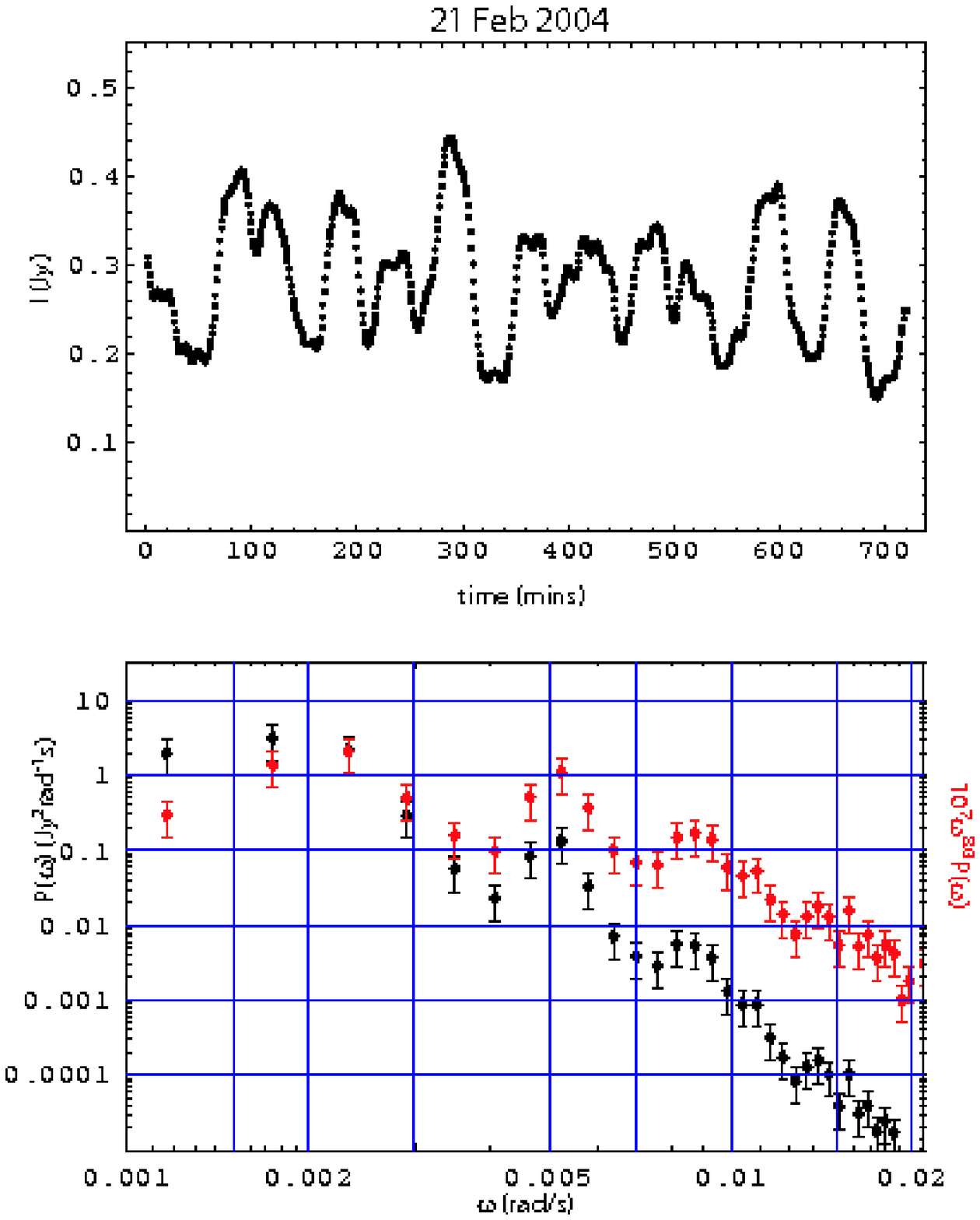,width=6.0cm} \\
 \psfig{file=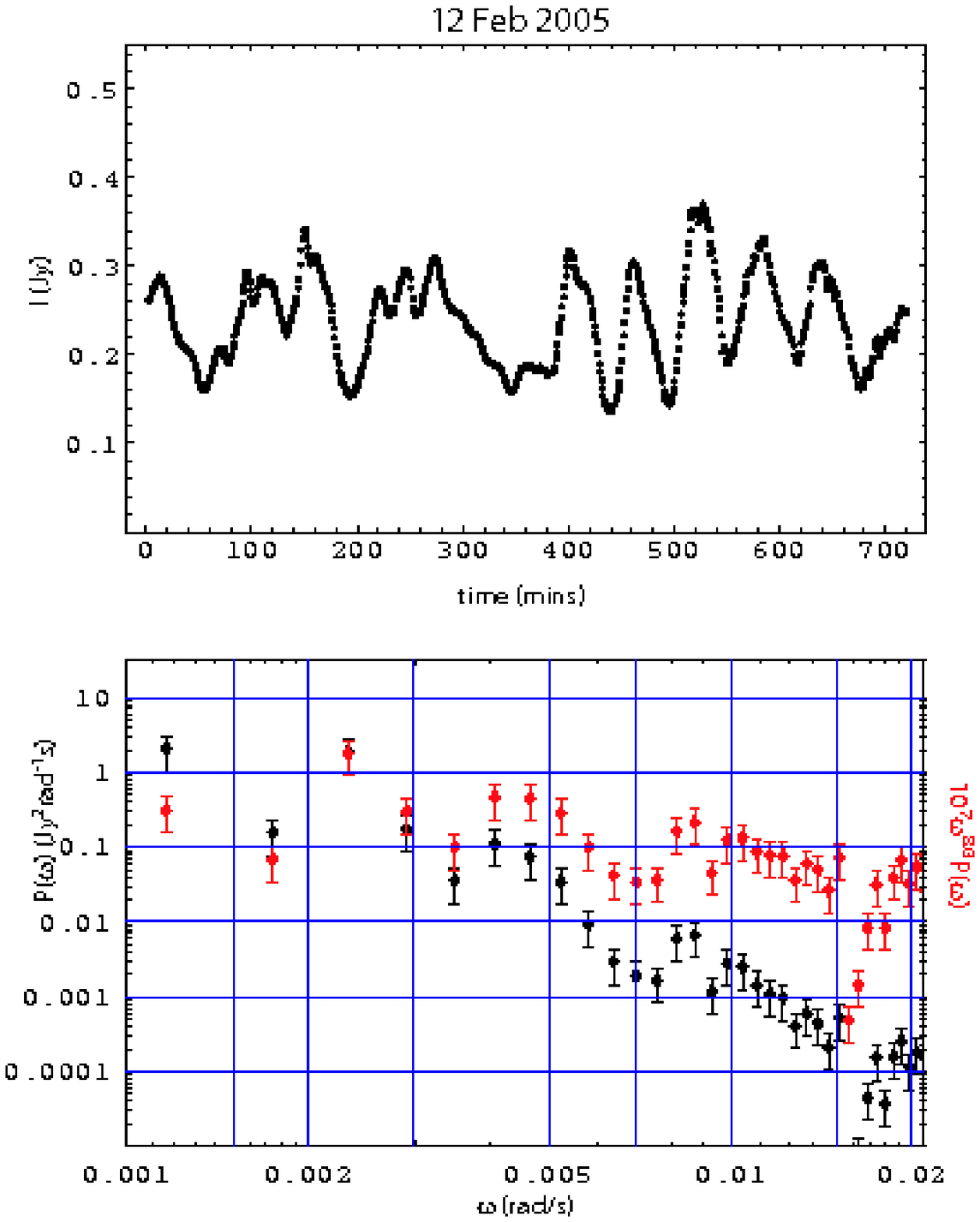,width=6.0cm}  &
 \psfig{file=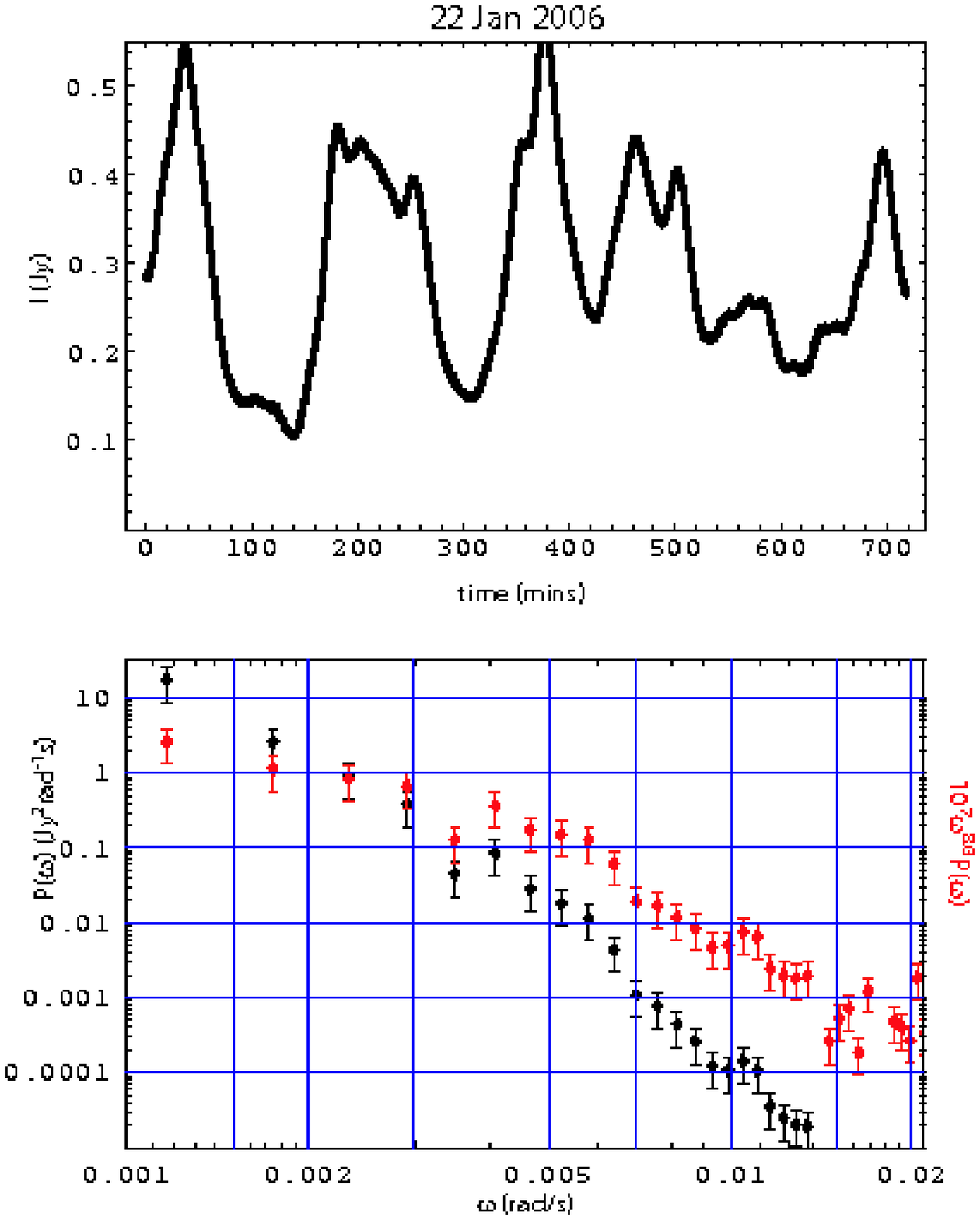,width=6.0cm} & 
 \psfig{file=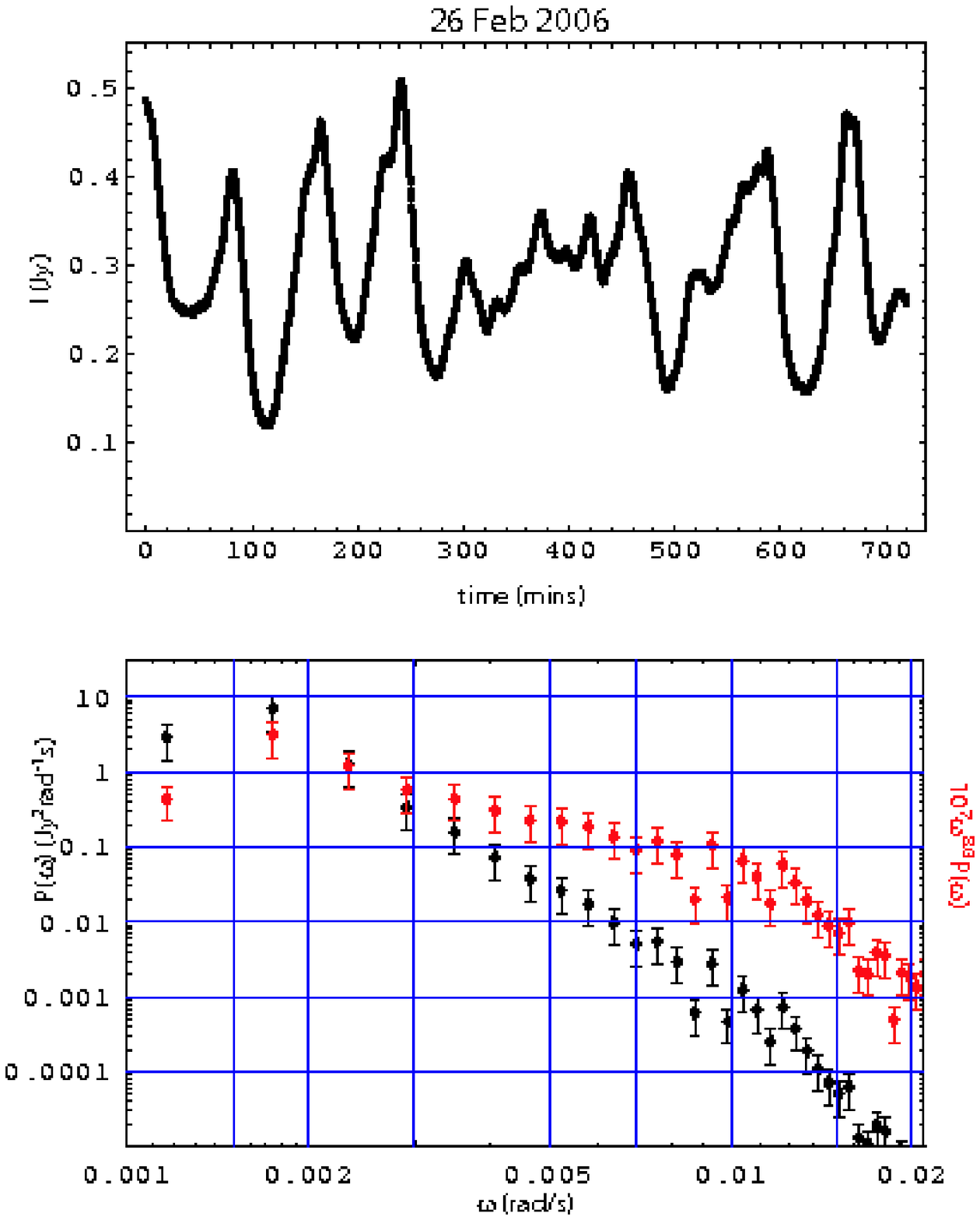,width=6.0cm} \\
 \end{tabular}
 \end{center}
  \caption{The $\mu$as structural evolution of J1819$+$3845, demonstrated by changes in character of the lightcurves and their associated spectra (black points).  The red points show the power spectra weighted by $\omega^{8/3}$ (see text for details).  The spectra were pre-whitened to prevent leakage of low-frequency power associated with the time sampling window function to high frequencies.  A noise bias, determined from the spectral power near 0.1\,rad\,s$^{-1}$, is also subtracted from the spectra (see Macquart \& de Bruyn 2006).  To reduce the errors, at the expense of information at low $\omega$, each spectrum shown is the average obtained by computing separate spectra for each 3-hour block of data within each 12-hour lightcurve.  The error on each point is half of its mean value and follows a $\chi^2$ distribution with 8 degrees of freedom.  It would be appropriate to normalise spectra by the source mean flux density if the entire source were subject to scintillation, but as this is not necessarily the case and it is misleading otherwise, the plots here are not so normalised.  The 2000-5 lightcurves were observed with 60\,s time resolution while the 2006 lightcurves were observed with 10\,s resolution.  The source mean flux was around 300\,mJy over 2003-6 after a fast rise from 125 to 300\,mJy between 2000 and 2003.  
} \label{LightcurvesFig}
\end{figure*}

Figure \ref{LightcurvesFig} shows the lightcurves from observations made 
over six epochs in the period 2000-6 at 4.8\,GHz with the Westerbork 
Synthesis Radio Telescope.  Details of the data reduction procedure are 
described in DB03.  Each observation was made at the same time each year 
so that the scintillation velocity vector is identical on all occasions; 
no lightcurves during the period 2001-2 are shown because we do not have 
good quality lightcurves near the pertinent dates.

The lightcurves exhibit an increase in the degree of small timescale 
structure in 2004-5 compared to those spanning the interval 2000-3.  The 
power spectra of the variations develop substantial structure at angular 
frequencies between 0.005 and 0.02 rad\,s$^{-1}$, and peak-to-peak 
variations shorter than 15\,mins (corresponding to structure on a 
timescale $\approx 4\,$mins in the structure function) are observed in the 
2004-5 lightcurves.  This corresponds to the highly significant $\omega 
\approx 0.005\,$rad\,s$^{-1}$ bumps evident in the 2004-5 power spectra. 
An additional feature at $\omega \approx 0.008\,$rad\,s$^{-1}$ is also 
evident, albeit at a low power, in the 2005 spectrum.  Both these bumps 
are either absent or less evident in the 2000-3 and 2006 spectra, 
although several do possess a significant `knee' near 
$0.004$\,rad\,s$^{-1}$.

\begin{figure}
\begin{center}
 \psfig{file=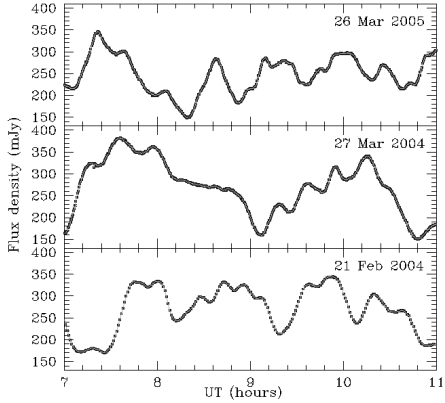,width=6cm}
 \end{center} 
\caption{Lightcurves of J1819$+$3845 at 4.8\,GHz taken during February and March 2004/5.  The Feb data have 60\,s sampling, the other two 30\,s.} \label{MarObsns}
\end{figure}

Analysis of similar power spectra taken on calibrator sources allow us to 
be confident that the power spectrum is not dominated by telescope noise 
and systematic effects down to angular frequencies $\omega \approx 
0.015\,$rad\,s$^{-1}$ (see Macquart \& de Bruyn 2006).  Another set of observations taken on 26-31 Mar 
over the period 2001-6 (see Fig.\,\ref{MarObsns} for a comparison between Mar 2004-5 variations with the 21 Feb 2004 variations) shows a similar recent 
increase in the amount of small scale structure in the lightcurves, but we 
shall concentrate on the analysis of the February datasets in this paper.

\begin{figure}
\begin{center}
\psfig{file=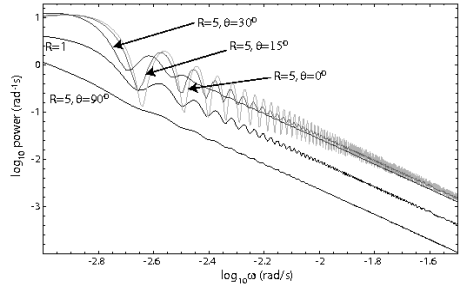,width=60mm}
\end{center}
\caption{Scintillation power spectra for a point source of unit flux density with $v_{\rm ISS}=50\,$km\,s$^{-1}$ and $z=10\,$pc for various $R$ and $\theta$.} \label{TheoryPowerSpectraFig}
\end{figure}

\section{Scintillation Modelling}

The differences between the character of the lightcurves over the interval 2000-6 can be interpreted either in 
terms of (i) changes in the interstellar turbulence responsible for the 
scintillations observed in J1819$+$3845, or (ii) changes in the internal 
structure of the quasar.  This is illustrated by the expression for the 
power spectrum of temporal variations in the regime of weak scintillation, 
applicable to the variations of this source at $\lambda\,6\,$cm (Codona \& 
Frehlich 1986;  Dennett-Thorpe \& de Bruyn 2000):
\begin{eqnarray}
P(\omega) =  \frac{1}{v_{\rm ISS}} \int_{-\infty}^{\infty} dq_y P_{\rm pt}\left(
\frac{\omega}{v_{\rm ISS}},q_y \right)  \left\vert V\left( \frac{ \omega  z}{v_{\rm ISS}k}, \frac{ q_y  z}{k}  \right) \right\vert^2, \label{Pomega}
\end{eqnarray}
where $V({\bf r})$ is the source visibility measured on a baseline ${\bf r}$, $k=2 \pi/\lambda$ is the 
wavenumber, ${\bf v}_{\rm ISS}$ is the scintillation velocity, here 
oriented along the $x$-axis, and
\begin{eqnarray}
P_{\rm pt}({\bf q}) = 8 \pi r_e^2 \lambda^2\Phi_{N_e} \left({\bf q}  \right)
 \sin^2 \left(  \frac{q^2\,z}{2k}\right)  ,  \label{Ppt}
\end{eqnarray}
is the spatial power spectrum of intensity scintillations that would be 
observed if the source were pointlike.  The amplitude of the line-of-sight-integrated electron density 
power spectrum $\Phi_{N_e}$, for turbulence located on a thin screen of thickness $\Delta 
L$ at a distance $z$, is ${\rm SM}=C_N^2 \Delta L$.  The scintillations 
are likely to emanate from only a single thin scattering screen because 
only a single screen velocity is required to model the annual cycle in the 
scintillation timescale (DB03).

Changes in the turbulence spectrum are disfavoured as the 
cause of changes in the variability characteristics because they require 
a rather contrived modification of the power spectrum over only a narrow spatial wavenumber  
range.  The appearance of spectral 
features over the frequency range $\omega \approx (4-10) \times 
10^{-3}\,$rad\,s$^{-1}$ requires an enhancement of power in the electron 
density power spectrum only over wavelengths $2 \pi v_{\rm ISS}/\omega 
\approx (3-8) \times 10^7 v_{50} \,$m (where $v_{50} \equiv v_{\rm 
ISS}/50\,$km\,s$^{-1}$).
Changes in the variability characteristics are more readily interpreted in 
terms of evolution in the micro-arcsecond source structure.  Such 
evolution is expected since both the intrinsic flux density and 
polarization are observed to change in this source on $\ga 0.5\,$yr 
timescales.

The $\sim\!15$\,min variations in the 2004-5 lightcurves require 
the presence of ultra-compact structure within the source.  However, the 
detailed shape of their power spectra depend on both the source structure 
and the scintillation physics.  The peaks observed in the 2004-5 power 
spectra could be due to either structure in the source visibility function 
or oscillations caused by the sine-squared term in eq.\,(\ref{Ppt}), the 
`Fresnel filter'.  The latter could become more prominent if the source 
becomes more compact.  Oscillations due to the Fresnel filter are 
exacerbated when the turbulence is anisotropic, here parameterized by 
introducing an elongation, $R$, in the electron density spectrum, oriented 
at an angle $\theta$ to ${\bf v}_{\rm ISS}$:
\begin{eqnarray}
\Phi_{N_e} = {\rm SM} \Big[ \frac{(q_x \cos \theta +q_y \sin \theta)^2}{R} 
 \nonumber 
+ R\, (q_x \sin \theta - q_y \cos \theta)^2 \Big]^{-{\beta \over 2}} . \label{PNeDefn}
\end{eqnarray}
Fig.\,\ref{TheoryPowerSpectraFig} shows the power spectrum for a 
point source for a variety of scattering geometries assuming the index for 
Kolmogorov turbulence of $\beta=11/3$.  Fresnel oscillations are 
potentially important in the interpretation of J1819$+$385 power spectra 
because its scintillation pattern is anisotropic.

It is difficult to attribute all the features of the 2004-5 power spectra  
to Fresnel oscillations because the spectral peaks change positions 
between years (see, for example, the 4\,mrad\,s$^{-1}$ feature in the 2004 spectrum).  
Even within a single power spectrum the locations of the peaks 
are inconsistent with a unique value of $r_{\rm F}$; numerical integration of eq.(\ref{Pomega}) shows that no combination of $r_{\rm F}$ and $\theta$ reproduces the peak locations observed in the 2004-5 spectra.
Nonetheless, detailed fitting described below suggests that the first peak, at $1.7$\,mrad\,s$^{-1}$, exhibited by many of the spectra, can indeed be 
attributed to the first peak of the Fresnel filter, and  
implies a screen distance $\sim 5 \, v_{50}\,$pc. 

The inability of the Fresnel filter to account for the peaks in the 2004-5 
spectra leads us to consider the additional influence of source structure.  
We first discuss simple arguments to deduce general source properties from 
the power spectra before reproducing them quantitatively using a specific 
source model.

We examine the size of the structure implied by the fast variations 
evident in the 2004-5 lightcurves. The scintillation power expected for a 
point source declines as $\omega^{-8/3}$ for $\omega \ga 2.5 \times 
10^{-3}\,$rad\,s$^{-1}$ and $\beta=11/3$ (viz. eq.\,(\ref{Pomega}) and 
Fig.\,\ref{TheoryPowerSpectraFig}).  For a point-like source the 
visibility, $V({\bf r})$, and hence $\omega^{8/3} \,P(\omega)$, would be 
flat.  Consider the $\omega^{8/3}$-weighted power spectra shown in red in 
Fig.\,\ref{LightcurvesFig}.  Fluctuations at $\omega$ depend on $|V(r 
\approx \omega r_{\rm F}^2/v_{\rm ISS})|$ so the frequency at which the 
red points depart from a flat line marks the point at which the source 
structure dominates the shape of the power spectrum.  This yields a {\it rough} estimate of the source size.
A sharp decline in $\omega^{8/3} P(\omega)$ at a scale $\omega_0$ indicates the presence of 
structure down to an approximate scale
\begin{eqnarray} 
\Theta 
\sim 3.3 \left( \frac{\omega_0}{0.01\,{\rm rad\,s}^{-1}} \right)^{-1} 
\left( \frac{v_{\rm ISS}}{50{\rm km\,s}^{-1}} \right) \left( \frac{z}{10{\rm pc}} \right)^{-1} \mu{\rm as}.
\end{eqnarray}
For instance, the 0.01 rad\,s$^{-1}$ break in the 26 Feb 2006 power indicates structure down to $\sim 7 \,v_{50} \, 
z_{5}^{-1}\,\mu{\rm as}$ if the index of the turbulence spectrum is $\beta=11/3$.  If the turbulence spectrum were shallower there is less source structure at high angular resolution, and if steeper there is 
more.

Since the bumps in the power spectra cannot be reproduced in terms of 
Fresnel oscillations, we consider the simplest source model capable of 
introducing additional spectral peaks.  A double source, 
with component flux densities $I_1$ and $I_2$ and sizes $\theta_1$ and 
$\theta_2$, with brightness power spectrum
\begin{eqnarray}
\left\vert V({\bf r}) \right\vert^2   &=&  I_1^2 e^{-k^2 r^2 \theta_1^2} + I_2^2 e^{-k^2 r^2 \theta_2^2}  \nonumber \\ 
&\null& + 
2 I_1 I_2 e^{-k^2 r^2 (\theta_1^2+\theta_2^2)/2}   \cos (k \Delta \btheta \cdot {\bf r}), \label{DoubleSrcEq}
\end{eqnarray}
induces oscillations in the scintillation power spectrum whose spacing 
depends on the component separation, $\Delta {\btheta} = \Delta \theta \, 
(\cos \alpha,\sin \alpha)$, where $\alpha$ is the angle the vector makes 
with ${\bf v}_{\rm ISS}$.  The oscillations are partially damped if the mean square of the component angular sizes exceeds $\sim \Delta \theta$.  

We performed fits to two power spectra up to $\omega = 
0.015$\,rad\,s$^{-1}$, the point beyond which the power dips below the
level at which we are able to confidently measure flux density variations. 
The 26 Feb 2006 spectrum, closely resembling the 
2000-3 spectra, was used as representative of spectra with the simpler 
structure, while the 21 Feb 2004 spectrum is representative of data in 
which additional fast, small amplitude variations are prominent.  The fit 
was performed to both spectra simultaneously to ensure the 
scattering parameters, ${\rm SM}, z, R, v_{\rm ISS}$ and $\theta$, were common 
to both.  

\begin{figure}
\begin{center}
\psfig{file=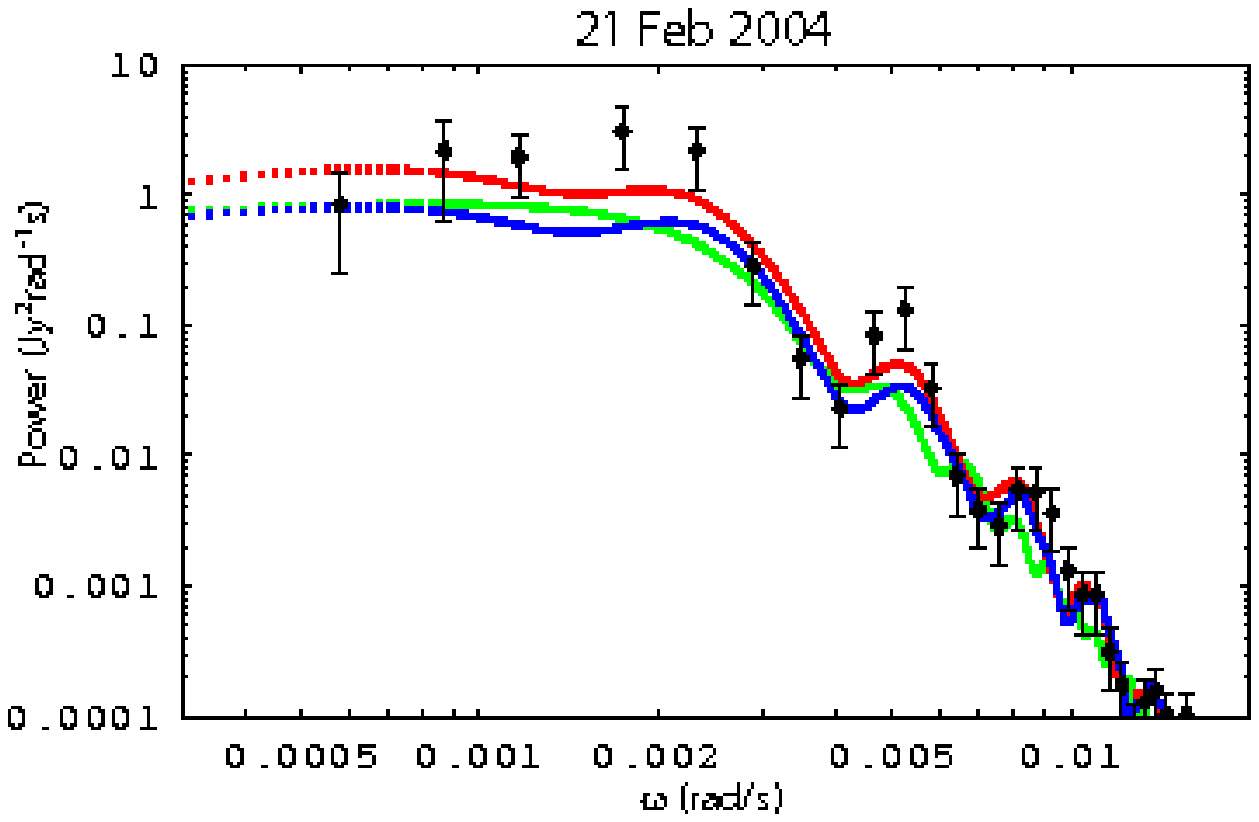,width=5.6cm} 
\psfig{file=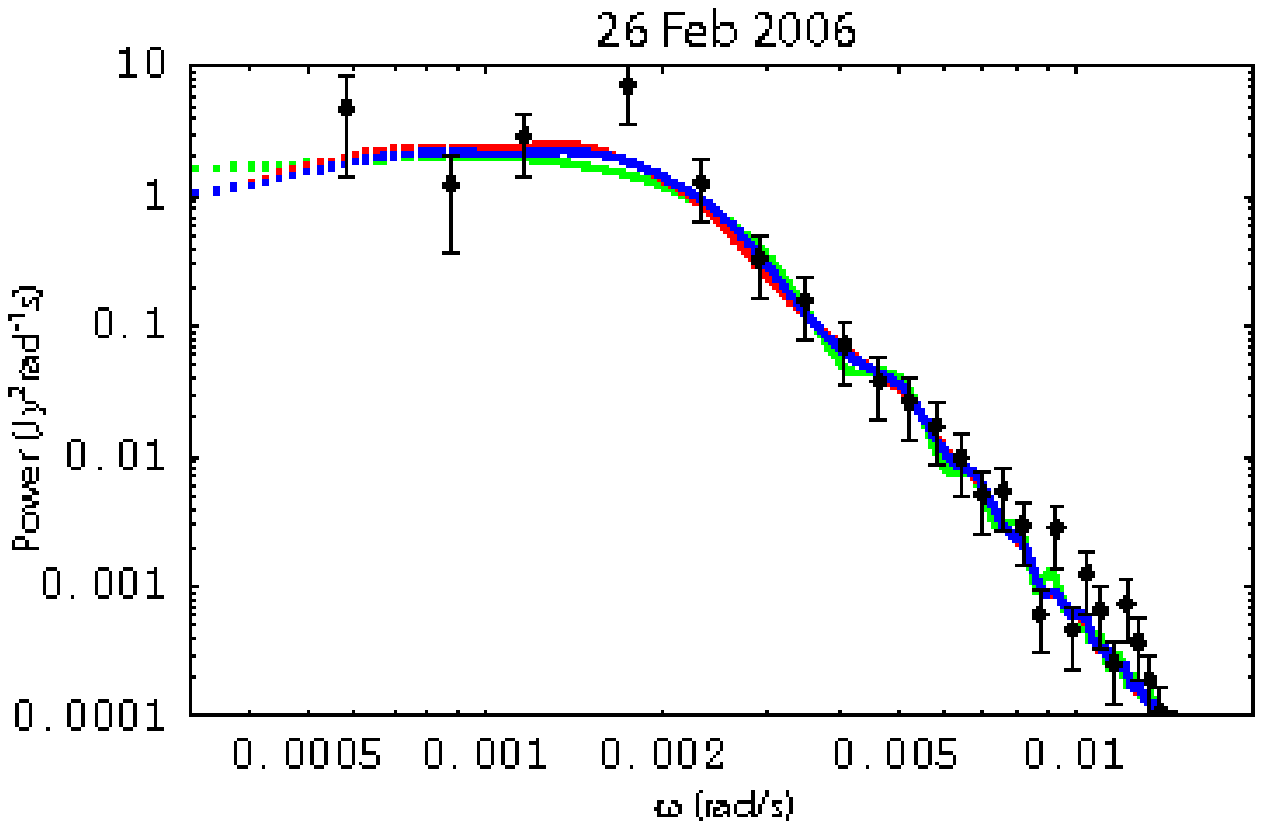,width=5.6cm} 
\end{center}
\caption{A fit to the 21 Feb 2004 and 26 Feb 2006 power spectra with the parameters of Table 1 (M1 blue line, M2 green line).  The red line represents M1 but with $R= 3.5$.  The higher anisotropy fit provides a better fit  to the low-frequency part of the 2004 spectrum, but a worse fit overall with a reduced $\chi^2$ of 1.3 vs. 1.0.}
\label{FitFig}
\end{figure}

The best-fitting parameters are listed in Table\,\ref{FitTable}.  Double 
component models were fit to both spectra.  Fig.\,\ref{FitFig} shows that the 
models reproduce the essential characteristics of the power spectra well. 
Two models, M1 and M2, were fitted for.  They differ in that $v_{\rm ISS}$ and $\alpha$ were fixed in M2 to values suggested by the annual cycle in the source's variability timescale (DB03).  In fixing $\alpha$ instead of $\theta$ we assume that the source structure, rather than anisotropy in the turbulent scattering plasma, dominates the anisotropy of the source's scintillation pattern.  This is justified by the low value of $R$ derived in all fits (see below).  However, we caution that the large anisotropy of the scintillation pattern renders the solutions for $v_{\rm ISS}$ and $\alpha$ degenerate in the annual cycle fit described in DB03.  Thus it is not clear that the ``best-fit'' parameters suggested in DB03 do indeed represent the correct solution.

The quality of the fit was degraded by fitting the spectra with a single source component.  However, we note that the best-fit flux density of the second component is only 2-3$\sigma$ from zero.  A fit without the second component in the 2006 data yields a reduced $\chi^2$ of 1.2.  The double-component model considered here is nonetheless strongly justified by polarization observations: time delays are observed between the polarized and unpolarized variations of this source over the entire period 2000-6, demonstrating that the source is always comprised of at least two components, one of which is likely highly polarized (Macquart, de Bruyn \& Dennett-Thorpe 2003).
There was no improvement in the fit quality by allowing $\beta$ to vary from its 
assumed value of $11/3$; the reduced $\chi^2$ for the fit was 1.02 and yielded $\beta=3.87$. 
However, other possible islands of low $\chi^2$ conceivably exist besides those detected by our fitting program and may yield acceptable fits with values of $\beta \sim 5$.

The $R=2.9\pm 1.0$ turbulence anisotropy derived from the fit is smaller than 
the $15_{-8}^{+>30}$ anisotropy of the scintillation pattern itself (DB03), but is not 
inconsistent because the latter is also influenced by asymmetry in the source 
structure. The value of $R$ is particularly sensitive to the low-frequency portion of the power spectrum. 
Two low-frequency points, obtained by recomputing the power spectrum from the lightcurves 
subdivided into 6\,hr blocks, were therefore added to the spectra employed in the fitting (see Fig.\ref{FitFig}).  This additional information is obtained at the expense of poorer error 
estimates associated with these points.  The derived anisotropy ratio should 
therefore be regarded with some caution; we expect to derive a better 
estimate in future from longer duration observations. 

There is a degeneracy between SM and the component flux densities; only the product ${\rm 
SM} \,I^2$ affects the spectral amplitude.  Since the sum of the 
component flux densities cannot exceed the 281 and 295\,mJy mean flux 
densities observed in the 2004 and 2006 lightcurves respectively, the highest possible values of $I_1$ and $I_2$ in the 2006 spectrum in M1 are 236\,mJy and 60\,mJy respectively, requiring ${\rm SM}>2.1 \times 10^{16}\,$m$^{-17/3}$.  This places a lower bound on 
the turbulent amplitude of $C_N^2> 0.68\, (\Delta L/1\,{\rm pc})^{-1}\,$m$^{-20/3}$.
For comparison, Macquart \& de Bruyn (2006) deduced ${\rm SM}=(1-1.4) \times 10^{17}$\,m$^{-17/3}$ for $z=4\,$pc in a fit to the refractive scintillation at 1.4\,GHz.
A Virginia Tech Spectral-Line survey measurement (see http://www.phys.vt.edu/$\sim$halpha/) of the H$\alpha$ intensity at the position of J1819$+$3845 implies ${\rm SM}=1.1 \times 10^{17} T_4^{0.9 } l_0^{-2/3} \epsilon^2/(1+\epsilon^2)\,$m$^{-17/3}$, where the gas temperature is $T=10^4 T_4\,$K, $l_0$ is the turbulence outer scale in AU and $\epsilon^2 = \langle (\delta n_e)^2 \rangle /\bar{n_e}^2$ is the normalised electron density variance.   This suggests $l_0 \la 9 \,$AU if $\epsilon \sim 1$ and $T_4 \la 10$.

A weaker degeneracy between the screen distance and component 
angular sizes allows decreases in $z$ accompanied by corresponding increases in 
all measured angular sizes to yield fits of {\it similar} quality.  However, changes in $z$ do adversely affect the fit because they also alter the positions of the Fresnel oscillations.

\begin{table}
\begin{tabular}{lllll}
\null & \null & \multicolumn{1}{c}{M1 } & \multicolumn{1}{c}{M2} & \null  \\
reduced $\chi^2$ & \multicolumn{2}{r}{$39.1/39$d.o.f. $=1.0$} & \multicolumn{2}{l}{$45.8/42$d.o.f. $= 1.1$} \\ 
Fit Parameter & \null & \null & \null & \null \\
\hline
SM\,($10^{17}$m$^{-5.67})\!\!\!\!\!\!\!\!$ & \null	&  $2.6 \pm 0.2$  & $2.4 \pm 0.2$ &\null  \\
$z$\,(pc)						&   \null   	&  $3.8 \pm  0.3$ & $2.0 \pm  0.3$ & \null 	\\
$v_{\rm ISS}$\,(km\,s$^{-1}$) 		&   \null 	& $59.0 \pm 0.5$  & $42.5^*$ &\null  \\
R							&   \null	&  $2.9 \pm 1.0$   & $1.9 \pm  0.6$ & \null  \\
$\theta$\,(rad)					&   \null	&  $2.26 \pm 0.02$	&$0.9 \pm  0.2$ & \null \\
\hline
\null & \multicolumn{2}{c}{2004} &\multicolumn{2}{c}{2006}  \\
\null & \multicolumn{1}{c}{M1} & \multicolumn{1}{c}{M2} & \multicolumn{1}{c}{M1} & \multicolumn{1}{c}{M2} \\
\hline
$I_1$ (mJy)					&   $42\pm7$ & $69\pm 30$		&  $67\pm 11$ & $100\pm45$	\\
$I_2$ (mJy)					&   $11\pm 4$ & $9\pm4$		& $17\pm 12$ & $25\pm13$	\\
$\Delta \theta$ ($\mu$as)			& $240\pm15$ & $220\pm20$		& $565\pm 15$ & $556\pm16$	\\
$\alpha$ (rad)					&   $0.00\pm 0.03$ & $1.07^*$	& $2.4\pm 0.3$	& $1.07^*$ \\
$\theta_1$ ($\mu$as) 			& $16 \pm 1$ & $18\pm2$	& $20\pm 3$ & $26\pm4$  \\
$\theta_2$  ($\mu$as) 			& $16\pm 2$& $0\pm500$		& $7\pm 4$ & $9\pm3$ \\
\end{tabular}
\caption{Fit parameters with 1-$\sigma$ errors for the 2004 and 2006 power spectra derived using the CERN MINUIT minimization package, which estimates errors based on the second derivatives of $\chi^2$.  
Fit parameters are defined in the text.  Fitting models M1 and M2 differ only in that parameters marked by asterisks are held fixed in M2.  The redshift-corrected brightness temperature of component 1 in M1 is $1\times 10^{13}\,$K.} \label{FitTable}
\end{table}

\section{Conclusion}

During 2004-5 fast, $\sim 15\,$min, variations emerged over the 
$\sim 30\,$min variations normally observed in the lightcurve of the 
scintillating quasar J1819$+$3845.  In 2006 the 
source reverted to variations similar to those in 2000-3.
The changes are best explained in terms of evolution in the source structure. A double-component source is capable of explaining the power spectra during 2004-5.  The spectral peak at $\omega \approx 2 \times 10^{-3}$\,rad\,s$^{-1}$ corresponding to the large amplitude, slow 
variations and visible in all spectra, is explained by oscillations of the 
`Fresnel filter', from which a $5\,v_{\rm 50}\,$pc distance to the 
scattering medium is deduced.  Holding the scintillation velocity fixed at the value derived by DB03 yields $z=2.0 \pm 0.3$, while fitting also for $v_{\rm ISS}$ yields $3.8\pm 0.3\,$pc.   The fast variations in 2004-5 are best attributed to oscillations in the source visibility function caused by a 
double-source structure with a component separation of $240\pm15\,\mu$as in 2004.  From a fit to the 2006 spectrum one infers a separation of $565\pm 15\,\mu$as. This implies an apparent expansion speed of $3.4 \pm 0.3 \,c$ over the two years separating the two observations. However, we caution that it is also possible to fit the 2006 spectrum with in a single component, in which case no expansion speed can be derived.

The emergence of these fast variations preceded other 
changes in the source, namely an increase in the modulation index and then the intrinsic flux density.  The appearance of new structure may be 
connected with the diffractive scintillation reported at 21\,cm (Macquart \& de Bruyn 2006), but as absolute astrometry is not possible using scintillation techniques one cannot determine the 
angular offset between the components responsible for 6 and 
21\,cm variations.

The proximity of the scattering screen and the large amplitude of the 
intensity variations requires the scattering turbulence to possess an amplitude $C_N^2 > 1.7\,{\rm m}^{-20/3}$ if 
localized to a region of thickness no more than 0.4\,pc, 10\% of its  distance.  This exceeds $C_N^2$ values deduced from most scattered pulsars (Cordes et al. 1988) by over two orders of magnitude.


\section*{Acknowledgments}

The WSRT is operated by the Netherlands Foundation for Research in 
Astronomy (NFRA/ASTRON) with financial support by the Netherlands 
Organization for Scientific Research (NWO).  JPM thanks Mike 
Ireland and Barney Rickett for suggestions regarding power spectra.
The Virginia Tech Spectral-Line Survey (VTSS) is supported by the National Science Foundation. 

\label{lastpage}

\end{document}